# Stochastic Flight Plan Optimization

Ítalo Romani de Oliveira, Steve Altus, Sergey Tiourine

The Boeing Company

Euclides C. Pinto Neto

IEEE Member

Alexandre Leite, Felipe C. F. de Azevedo

MWF Services

*Abstract* — Airlines have to achieve profitability while subject to highly uncertain factors, such as weather, varying demand, maintenance events, congestion, etc. The interplays among these factors are complex and large amounts of information are ignored due to difficulties in processing big data sets and finding useful correlations. This paper presents a method to better utilize existing sources of weather forecast and payload data for achieving more efficient flights.

The prediction of weather conditions has a high impact on flight efficiency, especially for long-haul flights. Despite there being many flight plan optimizers in the market, most of them rely on deterministic weather forecast, which entails the prediction of a single temporal series of weather conditions, but there are alternatives to this approach. We explore the use of stochastic weather forecast, which presents several distinct weather predictions for the same time horizon, forming a so-called ensemble forecast, where the value of a forecast parameter in each of ensemble member is different. So far, ensemble forecasts have been very little exploited for flight planning purposes.

Even without using ensemble forecasts, airlines carry out lots of statistical analyses based on past data, and devise effective policies for advising how much fuel and payload an aircraft should carry and how much of time buffer should be used in the schedule. But these buffers can be further reduced by doing forward-looking stochastic optimization. In this context, the use of ensemble forecast allows to select a trajectory that optimizes the expected outcome of a flight for an array of scenarios, instead of optimizing for a single one. Besides the weather, aircraft payload is another considerable source of uncertainty due to the lack of actual measurement of passenger and carry-on baggage weights.

We tested stochastic optimization, first with the objective of optimizing single flights, then with the objective of optimizing whole schedules. In one of the experiments for optimizing single flights, it was observed that, in 55.8% of the cases, stochastic optimization outperforms conventional optimization in terms of fuel consumption; in only 0.4% of the cases, conventional optimization wins; and, in the remaining 43.8% of the cases, they achieve equal results. The experiments with stochastic payload optimization demonstrated that the use of payload uncertainty can squeeze a bit more fuel savings from the flight plan outcomes. But the use of this technology is not driven only by reducing overall fuel consumption. One optimization criterion can be the minimization of diversions or fuel emergencies, that is, choosing the candidate that minimizes the maximum fuel consumption (minimax). In a project derived from this system, the optimization of whole flight schedules is performed, and its benefits were assessed.

## I. Introduction

Flight operations are an intrinsically stochastic activity, subject to multiple uncertainties, among which weather and traffic are the most significant ones. Predicting the amounts of fuel and time that a flight will take can be done only approximately and, because of that, there are safety regulations imposing reserve fuel rules. These rules masquerade prediction errors because their extra values are much larger in comparison with those errors. Although the regulations fulfill their safety purpose, the prediction errors remain as sources of inefficiency, influencing it in two ways: first, by the law that carrying fuel along the flight consumes more fuel, and second, by the trade-offs among payload, fuel and range. These laws influence the efficiency of flight operations and, by consequence, the profitability of an airline or flight operator.

Existing weather forecasting systems are highly sophisticated, but their output always contain errors, which are proportional to the advance time of the forecast. Estimations with longer time horizons are less accurate than those with a shorter window, and even these contain residual errors that cannot be eliminated. Some severe weather events such as storms are even less predictable than temperature and background winds. Even with these difficulties, the weather forecasting models can provide information about their uncertainties, which helps to improve flight trajectory optimization. An analogous situation happens for traffic and several other uncertainty factors. Any information helping to quantify uncertainties influencing flight performance can help leveraging optimization.

Stochastic Optimization [1], a.k.a. Stochastic Programming, distinguishes itself from deterministic optimization in the major aspect that the latter assumes perfect information about the input variables, while the former uses characterizations of variable uncertainties to optimize the *expected* value of the outcome over a stochastic distribution of outcomes. Even if a deterministic optimization algorithm uses accurate estimations of statistical moments of the input variables, such as their mean and standard deviation, its way of working and results are fundamentally different than that of stochastic optimization. The principle of stochastic optimization can be understood by a simple example. Suppose that one has a system for estimating certain flight route times. Today, for going from the origin location $A$ to the destination location $B$, this system estimates that route $R_1$ takes 90 minutes and that route $R_2$ takes 92 minutes. A deterministic optimizer assumes perfect information and chooses route $R_1$. On the other hand, a stochastic optimizer computes the travel time in different weather and traffic scenarios, according to the uncertain variables used in time estimations, and may conclude that is better to take $R_2$ because of a high probability that the actual time for $R_1$ be much larger, say 100 minutes, and the statistical expectation of $R_1$ trip time is 93 minutes. Thus, it apparently sacrifices today's choice to win in the long run.

The elements of stochasticity that we explore in this work are the weather forecast and payload, in order to develop a stochastic optimization system. Other uncertainty factors are not in the scope of the present paper.

## II. FLIGHT PLAN OPTIMIZER

Aircraft trajectory optimization has many constraints which can be summarized in the requirement that it must be legally flyable. Besides using an accurate model of the aircraft performance in the different flight conditions (take-off, climb, cruise, descent), it has to integrate this model in a plausible trajectory, taking into account the weather, and observing the airspace regulations, such as allowed airways, forbidden air sectors, etc. Within the performance management constraints and inputs, there are legally mandated rules for reserve fuel that have to be observed (e.g. [2]), besides runway length and other factors. In order to deal with all these constraints in a productive way, there are professional Flight Planning Systems (FPS), used by flight operators to perform their daily flight plans, and issue legally flyable flight plans in standardized data formats, which can be easily understood by the flight crew and, ideally, loaded automatically into the aircraft Flight Management System.

Most FPSs perform some sort of optimization, but trajectory optimization is a very hard problem in the computational sense, so their optimization algorithms use simplifying heuristic rules. Some common optimization heuristics are the following: first, find an optimal horizontal path, using published waypoints and assuming a guess cruise altitude for the wind values, then proceed to find the best vertical profile that fits to it. For the vertical profile, assume the arrival condition (with the legal reserve fuel) at the destination airport, and find a backward-built approximation of the descent profile, until reaching the optimum cruise altitude; then find a forward-built feasible trajectory to get there, using the aircraft fuel capacity, including a guess of the climb trajectory. Then, refine this initial feasible trajectory progressively by segment, backwards, until the first cruise segment, and recompute the climb approximation. When reaching this point, it may happen that there is more fuel in the tank than needed, or that too little payload capacity was used, or both. Then, eliminate the excesses and start a second round of sequential optimization in the same fashion, and so on, until no further optimization is observed. Despite this scheme being already complicated, it does not reach theoretical optimality because it is sequential and does not explore the full space of possible trajectories. Some alternative approaches expand the trajectory search space [3, 4], but so far have not been incorporated to a commercial FPS.

Boeing has developed a new optimizer, called FlitePlan Core®, which is used for advanced research projects and expands the trajectory search space in 4-dimensions, beyond the sequential approach above described, and performs fast computations by using dominance rules [5] to quickly discard large subsets of non-improving solutions. Besides that, it has a state-of-the-art modelling of modern airspace constraints such as the Route Availability Document (RAD) from Eurocontrol [6]. With regards to practical use, it offers the capability of being called as a regular process via command line and has a small memory footprint, thus facilitating parallel running with different inputs in a multi-core setting. This tool, shortly referred to as FPC, is used throughout this research work as the baseline tool for deterministic optimization on which the stochastic optimization scheme is built.

Most of the commercial flight planning systems nowadays perform deterministic optimization, this meaning that they assume deterministic values for all of their inputs, including the weather forecast. The first step that we made towards stochastic optimization is to use the ensemble weather forecast, as described below.

## III. ENSEMBLE WEATHER FORECASTS

An ensemble weather forecast [7] is a collection of forecasts produced by the same model for the same forecast time, each representing a possible version of the future, and produced with perturbed values of the initial conditions, which are uncertain. These perturbations are generated according to a way of representing the probability distribution of the initial conditions, which simulates randomness by using certain deterministic mathematical formulations [8, 9]. The pioneer ensemble system was developed by the European Centre for Mid-term Weather Forecast (ECMWF), however we use the Global Ensemble Forecast System (GEFS) developed and maintained by the National Centers for Environment Prediction (NCEP) from the US [10], because it being freely available to the public.

GEFS provides 21 ensemble members in each issuance, from which one of them is the control member, produced with unperturbed initial conditions, and considered the "deterministic" forecast. Each member contains the same set of a few hundred atmospheric parameters, among which the most important for flight planning is the temperature and the 3-dimensional wind components. Currently there are two options of horizontal resolution for each ensemble member in the set, of 0.25 and 0.5 geodesic degree. The experiments performed by us used an older version with 1 degree of horizontal resolution. In the vertical dimensional, the GEFS forecast has 10 isobaric altitude layers. In the temporal dimension, it has a resolution of 6 hours, that is, there is one spatial grid for $t + 0$, $t + 6$, $t + 12$ and so on, up to 384 hours (16 days). Each GEFS forecast member covers the entire globe and is provided in the so-called GRIB format (General Regularly-distributed Information in Binary form), and an ensemble set is issued each 6 hours at the regular times of 00:00, 06:00, 12:00 and 18:00 UTC.

Each ensemble set issuance from GEFS has several Gigabytes and we had to store several months of these issuances on a local server, both for quick read access and because NCEP does not maintain a reliable store of the historical issuances.

## IV. EVALUATING THE PREDICTIVE POWER OF ENSEMBLE WEATHER FORECASTS

A fundamental question raised in the beginning of this research was if ensemble weather forecasts really had potential to increase the accuracy of predictions of flight fuel consumption and total cost, otherwise using them for stochastic optimization could show up savings that were false or unreliable due to larger errors in prediction. On the other hand, a mere increase in the accuracy of fuel consumption can help to increase the efficiency of flight operations. We developed experiments that indicated that indeed it was the case that ensemble weather forecasts allow higher prediction accuracy.

In order to probe that, we did not engage resources to collect all the data associated to real flights, however we used the weather reanalysis reports from the Global Data Assimilation System (GDAS) [11] from NCEP, which is a 3-D grid of weather data produced from the interpolation of many weather measurement reports around the world. The GDAS weather is the global weather data which most approximates the actual, present measurements, thus we used it as the ground truth for calculating the true flight costs. This data source can be also called weather *nowcast*.

The total flight cost is actually a summation of many terms, however in this context it is a weighted sum of the cost of fuel and the cost of time, where the weight of cost of time is given by the so-called cost index [12] or CI, in short. The weather influence on these factors is easy to grasp: flying with headwinds will either increase the flight time or, if that is compensated by increasing airspeed, that will require more thrust and thus more fuel. The cost index can be customized for each flight, because fuel is obviously not the only cost factor for a flight. Flying at the most fuel-efficient airspeed can be too slow for the passengers, require more crew hours and more aircraft availability, besides other operational considerations. Thus, throughout this paper, when we refer to flight cost, the value of the flight cost is computed according to the cost index methodology, with the CI weight variable assuming a single value for all instances in each sample, the exact values not being relevant for the analyses here. The CI-based flight cost is preferably expressed in units of mass of fuel (kg), in which the cost of time is implicit by converting a dollar amount to an equivalent fuel amount. Assuming that the fuel consumed is the predominant term in the CI equation, this choice of representation has a better filter for the fuel price variations and inflation.

In order to make comparisons between the flight costs computed with different weather data types, we start with the following definitions:

$C_D$: flight cost computed with deterministic weather forecast
$C_S$: flight cost computed with ensemble (stochastic) weather forecast
$C_A$: flight cost computed with the weather nowcast data, or ground truth flight cost.

The flight costs above are computed with FPC, by inputting to it the origin and destination airports, intermediate waypoints, and the weather forecast or nowcast. The ensemble weather data is composed by several ensemble members, thus a straightforward way to compute a single value for the flight cost is to run FPC with each ensemble member, obtain the flight cost for each one and take their average, according to

$$C_S = \frac{1}{N_S}\sum_{i=1}^{N_S} C_S^i, \quad (\text{Eq. 1})$$

where $N_S$ is the number of stochastic ensemble members, and $C_S^i$ is the flight cost computed with the $i$-th stochastic ensemble member. And, regardless of the method for computing $C_S$, the prediction errors can be computed as:

$$\epsilon_D = C_A - C_D \quad (\text{Eq. 2})$$
$$\epsilon_S = C_A - C_S \quad (\text{Eq. 3})$$

However, we have not sufficient knowledge on the ensemble principles to ascertain that taking the average of the costs for $C_S$ is the best estimation of the actual flight cost. Thus, we delegated this estimation to supervised Machine Learning (ML) regression algorithms, as per the scheme in Fig. 1. These algorithms take tuples $(C_A, C_D)_i$ and $(C_A, C_S^1, \ldots, C_S^{N_S})_j$, respectively, during training, and build predictor functions $\hat{C}_A = F_D(C_D)$ and $\hat{C}_A = F_S(C_S^1, \ldots, C_S^{N_S})$, respecively. Then the prediction errors become:

$$\epsilon_D = C_A - F_D(C_D) \quad (\text{Eq. 4})$$
$$\epsilon_S = C_A - F_S(C_S^1, \ldots, C_S^{N_S}) \quad (\text{Eq. 5})$$

This arrangement is illustrated by means of Fig. 1, where we included a third regressor $F_{D+S}(C_D, C_S^1, \ldots, C_S^{N_S})$, which takes into account the deterministic and stochastic forecasts together.

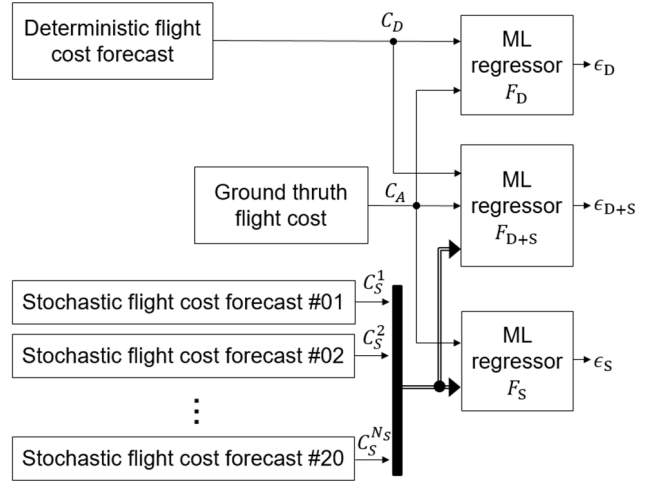

Fig. 1: ML regressors for increasing the accuracy of flight cost predictions.

In our experiments, the forecasts were issued 10 days before the nowcast day, in which the flight is supposed to happen. And we actually used several ML algorithms for the regressors, including: Deep Learning (or Multi-layer perceptrons), Random Forest, Gradient Boosting and XG Boost. As baseline for comparisons we used Equations 1-3, and the results of this comparisons can be summarized as in Fig. 2. The errors of the ML regressors were calculated with 10-fold cross-validation in a sample of 3,000 flights.

In this figure, it is interesting to see that the baseline (purple line) deterministic prediction outperforms all the ML regressors, and that the baseline stochastic prediction is just slightly better than the deterministic one. However, most of the ML regressors with stochastic forecast outperform the baseline prediction of Eq. 3. And the joint use of the deterministic and stochastic forecasts (D+S) does not improve performance, except for the Deep Learning regressor. Overall, by looking at these plots, it can be defended that the use of stochastic ensemble forecast has a significant potential for increasing the accuracy of flight cost predictions. Thus, with regards to prediction accuracy, there is no objection to use its data both for prediction and for consequent methods of optimization.

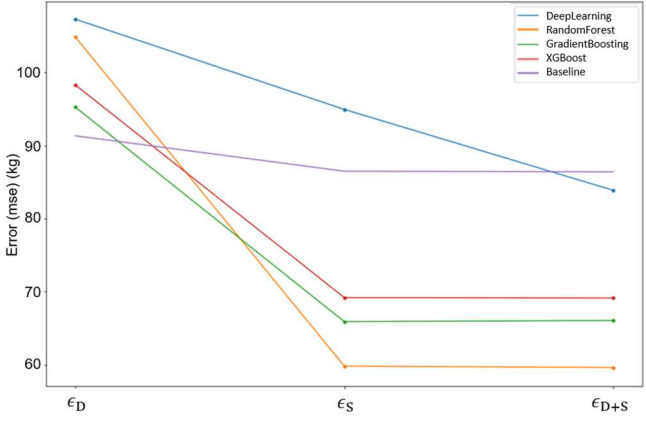

Fig. 2: Results of comparisons between ML regressors using deterministic and stochastic ensemble weather forecasts.

## V. Initial Version of Stochastic Optimization

We are mostly concerned with the optimization of a flight plan before the flight departs, because then it is possible to manage the amounts of fuel and payload that it will carry. Having an ensemble forecast with many versions of the future, weather, we seek a flight plan which optimizes a certain cost functional which is statistically defined with regards to the universe of possible outcomes of that flight. In practice, this is implemented according to the sketch of Fig. 3.

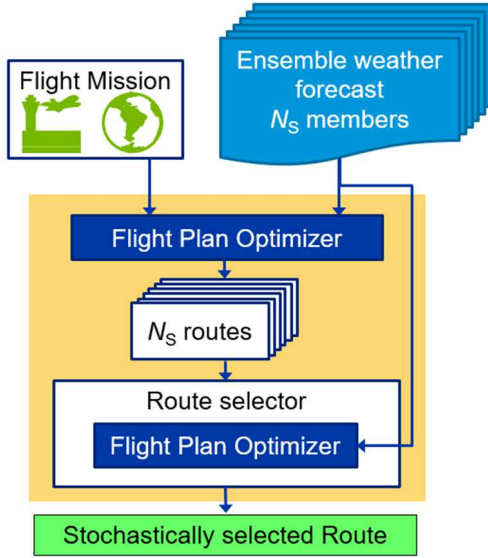

Fig. 3: Basic stochastic optimization method.

The flight plan optimizer generates potentially one route for each of the $N_S$ ensemble weather forecast members, although some routes may coincide due to the use of common waypoints and altitude levels. What really differs stochastic optimization from the deterministic one is to select a route from this set of routes according to some criterion, thus the box "Route selector" in Fig. 3. The route selection criterion can assume the following forms:
1) Minimize the *expected value* of the fuel consumed or the total flight cost.
2) Minimize the *maximum possible* fuel consumed.

Criterion 1 is the one that most appeals from the economic and environmental stand points, and the most usually applied in business optimization problems. Criterion 2 is more typical of game theoretic analysis, having the nickname *minimax*, but can be useful in practice for certain flights where a possible diversion (not landing at destination), due to a so-called fuel contingency, would have an extreme impact. Formulations according to Criterion 1 might have embedded the cost of diversion in the total flight cost, conditional on its occurrence, but that may not be enough in certain cases. Our research has focused primarily on using Criterion 1 for route selection.

It is important to note that route selection requires re-running the optimizer with the constraints determined by a route obtained in the first pass of the optimizer, that is, the route belonging to "$N_S$ routes" illustrated in Fig. 3. These routes are flyable flight plans and not strictly defined trajectories. Each of them may be flown in many different ways, and the factor which most varies in practice is the vertical profile. There is not much certainty on how will be the exact profile until the exact fuel and payloads are defined, and traffic deconfliction is performed. For practical purposes, at this point it is sufficient to have the sequence of waypoints as constraints for the next round of optimization. In our implementation with Boeing FPC as the flight plan optimizer, the explicit route constraints for the route selection phase are almost exclusively horizontal, namely waypoints and standard departure and arrival procedures (SIDs and STARs, respectively). Vertical constraints are implicitly defined in SIDs and STARs and in the other models of airspace structures and constraints internally used by FPC.

Using Criterion 1 for route selection means to use the following formula:

$$R^* = \arg\min_{i=1,\ldots,N_S} \left( \frac{1}{N_S} \sum_{i=1}^{N_S} C(R_i) \right) \qquad \text{(Eq. 6)}$$

Which consists of selecting the index $i$ corresponding to the route $R_i$ which minimizes the expected value of the cost $C(.)$ calculated in the second optimization pass. It is important to note that this second pass requires running the optimizer $N_S \times N_S$ times, that is, running each of the $N_S$ routes with each of the $N_S$ ensemble weather members. This can become a little clearer with help of Fig. 4.

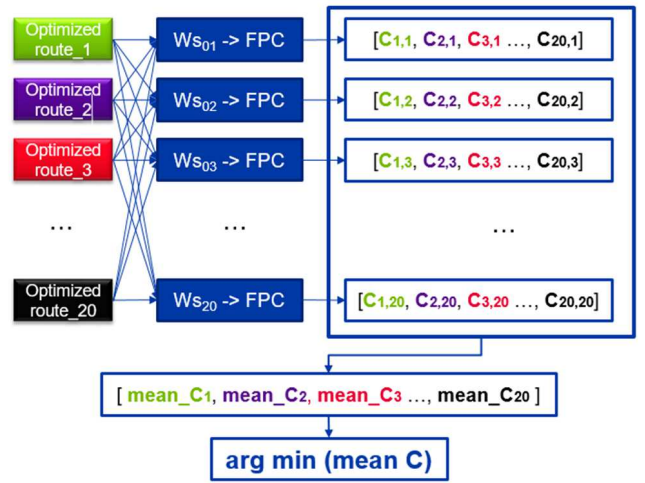

Fig. 4: Sketch of stochastic route selection with second-pass optimization.

In this figure we can visualize that each route to the left, that was optimized in the first optimization pass, is re-optimized by FPC with each stochastic weather ensemble member ($Ws_i$), and this generates a row of 20 cost values for each of these members, thus potentially 400 executions and corresponding cost values. The selection happens by taking the expected values of each of those columns, indicated by mean_$C_i$, and choosing the column with minimum value. As each column corresponds to each of the routes, as indicated by the colors, the best route is selected by this association.

A first validation of this method was carried out with the following definitions:

As origin-destination of the flights, we selected 12 round-trip city pairs with the following distribution:
- 4 city pairs between South America (SA) and North America (NA);
- 4 city pairs between SA and Europe (EU);
- 4 city pairs between NA and EU.

The dates the flights covered a period between August 2019 and May 2020 (~ 10 months), which was the period covered by our database of weather data. A total of 699 flight plans was used. The number of FPC runs for each city pair is accounted according to the following: 1 for the control member of the ensemble and 20 for the stochastic members, in the first optimization pass, thus amounting to 21 executions; each of the resulting routes is re-optimized in the second pass with each stochastic ensemble member, thus $21 \times 20 = 420$ in the second optimization pass. Thus, each city pair amounts to $420 + 21 = 441$ FPC runs per city pair. In our high performance virtual machine, this amounts to a couple of minutes of computing time.

However, we want to compare the performance of these optimizations with the deterministic optimization and all referring to a baseline cost, which we compute using GDAS [11], which adds 21 more FPC executions for each city pair. The deterministic optimization is already computed with the control member of GEFS [10], however in order to be fair we re-optimize the deterministic route from the first pass in the second pass, thus adding one more FPC execution. Thus, the whole comparison experiment required $441 + 21 + 1 = 463$ FPC runs for each of the 699 flights, amounting to a total of $463 \times 699 = 323{,}637$ FPC runs, corresponding to several hours of computing. We soon noticed that there was a significant rate of repetition in the routes generated in the first pass, then we developed a filter to catch that and skip repeated runs, which slashed the number of executions by about one half. The results of these comparisons are summarized below.

First, we wanted to know if the stochastic optimization actually dominates by producing more economical flights than the deterministic optimization. This summarized comparison is given in Fig. 5, which uses the fuel consumption as criterion.

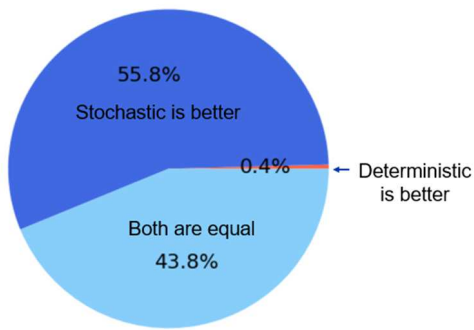

Fig. 5: Overall comparison between the first version of stochastic optimization and deterministic optimization.

As it can be seen in the figure, stochastic optimization clearly tends to produce more economical flights, using the GDAS-based flight cost as ground truth for a route. In 99.6% of the cases, stochastic optimization is capable of achieving the best result, and in only 0.4% the deterministic solution wins.

The size of the benefits achieved by stochastic optimization are shown below, broken down by the cost components, fuel and time. The plot of Fig. 6 shows the Probability Density Function (PDF) of the fuel savings obtained with flight plans from the stochastic optimization method in relation to the fuel consumption of the deterministic solution. That shows an average value of -33.3 kg, with a distribution skewed to the left and with a significant frequency of extreme savings corresponding to < -200 kg.

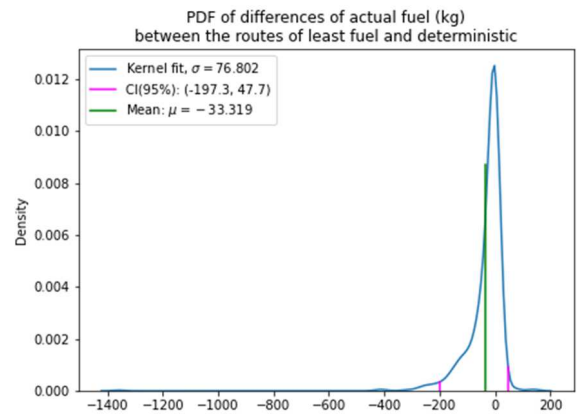

Fig. 6: Probability Density Function of fuel savings achieved by stochastic optimization.

The savings of flight time are shown in Fig. 7 in an analogous form. The shape of the distribution is similar to that of the fuel, and here the average time saving is only 0.164 minute, indicating the higher priority given in the CI for fuel consumption.

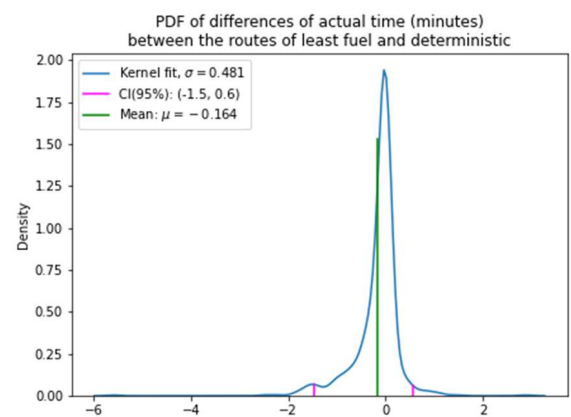

Fig. 7: Probability Density Function of time savings achieved by stochastic optimization.

Although the benefit of 33 kg of fuel for a long-haul flight does not sound a large one, when it is summed up for thousands of flights in an entire year, the figures start to have significance. One of Boeing's development partners tested out this initial version a couple of actual flights and provided a positive feedback, the experiment having been publicized in the Internet in the page [13], in the paragraph "Probabilistic Flight Planning".

## VI. OPTIMIZING ON PAYLOAD UNCERTAINTY

Payload uncertainty is another plague that contributes to decrease flight efficiency, both on the revenue side and on the operational side. Depending on the aircraft type used for a mission, it cannot carry its full payload due to meeting the required range with all legal margins of fuel. The aircraft operator makes an estimate of the available payload, considering the worst case scenarios and the possibility of diversion. The allowed payload with these considerations is what can be sold in advance, either as passenger seats or cargo. However, as the time of flight approaches, the worst case scenario can be re-assessed and become more accurate, especially with the use of ensemble weather forecast. At departure time, it is possible to know with a high accuracy the weight of the dispatched baggage and cargo, which is measured precised with the weighers. The weight of catering material can be known to a certain accuracy. However, weighing human beings and their carry-on is a great inconvenience, by many considered a privacy violation, and in most of the countries, people have the right to refuse being weighed. Thus, weighing passengers and their carry-on is not practiced by most airlines. Instead, they have methodologies to estimate passenger weight based on statistics. But this embodies a significant uncertainty in the total aircraft weight, that may play a role in flight efficiency and in the ticket price.

Using data from the US Bureau of Statistics, a rough figure for the an adult's standard deviation of body mass is $\sigma_p = 21$ kg, considering an average between both genders. Without considering carry-on baggage, the standard deviation of the sum of body masses of passengers in an airplane is:

$$\sigma_P = \sqrt{\sigma_p^2 \times N_P}, \quad \text{(Eq. 7)}$$

where $N_p$ is the number of passengers and $\sigma_P$ has an uppercase $P$ to be distinguished from the lowercase $\sigma_p$, and $\sigma_P$ becomes the total payload standard deviation. However, in practice, there are other uncertainties in the total mass uncertainty and it is better to use more comprehensive data from airline records to estimate that. From a sample database that we used, we obtained a figure of $\sigma_P \approx 5\%$ of the total payload mass. In our experiments, we used the latter alternative for the database of flights that we had.

The addition of this single source of uncertainty required us to develop a more elaborate flow of computations, and this made the number of FPC calls to jump enormously. The new flow of computations is illustrated in Fig. 9. From the probability distribution function of passenger payload, assuming a normal distribution with a certain mean and standard deviation $\sigma_P$, 5 equally representative values of payload are generated. For each of these values, all the weather ensemble members are input separately to FPC, thus requiring $5 \times 21 = 105$ FPC calls. For each different output route from these runs, the second optimization pass is executed, re-applying all combinations of weather and payload, thus requiring potentially $105 \times 105 = 11,025$ FPC calls. This means having one table of cost values with 105 entries for each of the 105 possible weather-payload combinations, each entry represented in Fig. 8 by the notation $f_{i,j}$, as a reference to the objective function of optimization, which is the flight cost. Each of these tables are averaged, and the route corresponding to the table with the minimum flight cost is selected as the optimum one. As before, the actual number of different routes is reduced significantly when eliminating repeated cases, however the computational load is still formidable.

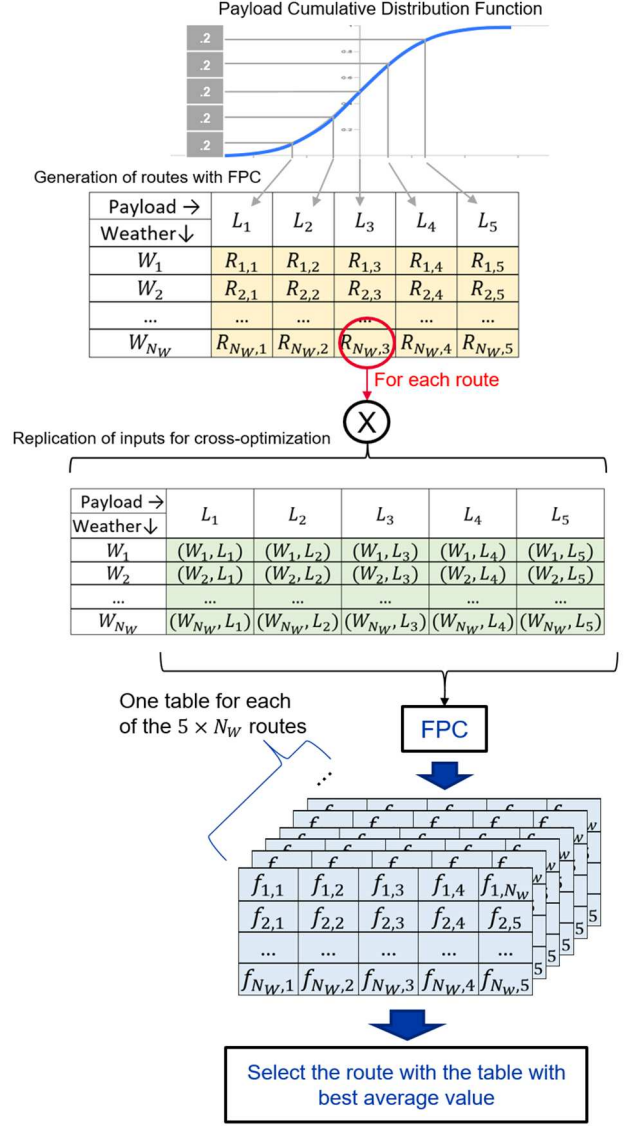

Fig. 9: Computation flow of stochastic optimization with uncertain payload.

In such context, it was not worth running a large number of flights for assessing the added optimization benefits, thus the new sample of flights consisted of 29 city pairs around the world. Out of the 29 flights optimized with this method, 11 obtained optimal routes different from the route obtained with the method of Section V, which optimizes only on weather uncertainty. Fig. 10 illustrates the results of this experiment for the 11 flights with different optimum route. The origin and destination airports (O_D) are discriminated by IATA code [14]. By looking at those results, they seem to confirm that the uncertain payload optimization had an effect, although the statistic indicators T-test S statistic and p-value (evaluated with the 29 flights) say otherwise, that the differences are not statistically significant. Even so, the fact that only 1 out of the 29 values had a worse value, and 10 out of 29 had a better value, seem to indicate that there is some latent potential on payload optimization. The fact is that there are other sources of payload uncertainty than passenger and carry-on body mass. Even the empty aircraft mass itself varies with time, by dirt accumulation, water retention, painting and paint decay, etc. Measuring actual aircraft weight is a technically challenging task and can be done in just a few specialized places, thus an aircraft cannot be weighed often throughout its useful life. Although this certainly contributes to increase mass uncertainty, we do not have reliable quantitative data to augment our calculations.

| Flight O_D | Fuel consumption (kg) with fixed payload | Fuel consumption (kg) with uncertain payload | Difference (kg) |
|---|---|---|---|
| AMS_YYZ | 56,448 | 56,414 | 34.25 |
| GIG_MIA | 60,902 | 60,905 | -3.48 |
| GRU_DFW | 75,421 | 75,428 | -1.72 |
| BCN_GRU | 83,533 | 83,539 | -5.97 |
| FRA_MCO | 73,219 | 73,266 | -47.49 |
| MIA_GRU | 57,658 | 57,745 | -86.68 |
| YYZ_LHR | 45,138 | 45,140 | -1.5 |
| CDG_YUL | 53,944 | 53,961 | -16.33 |
| JFK_AMS | 46,664 | 46,674 | -9.71 |
| MEL_LAX | 113,865 | 113,877 | -12.89 |
| SYD_LAX | 107,828 | 107,838 | -10.78 |
| | | Mean /11 | -14.75 kg |
| | Weighted mean percentage / 11 | | -0.021% |
| | | Mean /29 | -5.59 kg |
| T-test S statistic: | -1.6 | T-test p-value: | 0.138 |
| Positive values | 1 | Negative values: | 10 |

Fig. 10: Stochastic optimization results per flight compared, between the cases with fixed payload and with uncertain payload.

## VII. CONSIDERATIONS FOR PRACTICAL USE

The most probable users of this optimization system will be airline dispatchers, and one of key criteria for adoption of the tool is if these users will trust in its results. As mentioned earlier, using a single optimization objective function may not be desirable, and proposing counter-intuitive routes may look suspicious. The system may incur in routes that occasionally yield worse results, and that could scrap user confidence. Experienced airline dispatchers actually have an extensive knowledge of the factors in play, among which many are not captured by an optimization tool. Thus it is important to count on the dispatcher decision making skills and, for that aim, we developed some ways for the user interact with the stochastic optimizer in a more participative way.

One way of doing a productive interaction with the dispatcher is to show the routes on a map and their corresponding fuel or cost attributes in a boxplot diagram, and ordering them according to some criterion, as shown in Fig. 11. On top the routes generated with the ensemble weather forecast members are shown, between Charleston, SC, and Abu Dhabi; and on the bottom their corresponding fuel consumption boxplots, showing the mean values interquartile ranges, their standard minima and maxima, and outliers. That box plot is ordered by ascending mean fuel consumption (its expected value). Thus, using mean fuel consumption as criterion, the selected route would be s5. However, as indicated in the overlay, route s2 is the most robust against fuel contingency, being the minimax case for fuel consumption, despite having a higher expected fuel consumption than that of s5. By looking at the map, a dispatcher might quickly discard any route that does not meet some geographical criterion, any other operational reason.

This is just a simplified exhibit of how that interaction may play out. We have developed a simple working prototype with these elements, but putting that in a figure would be too cluttered. The key point is to keep the human decision makers in the loop and leverage their expertise.

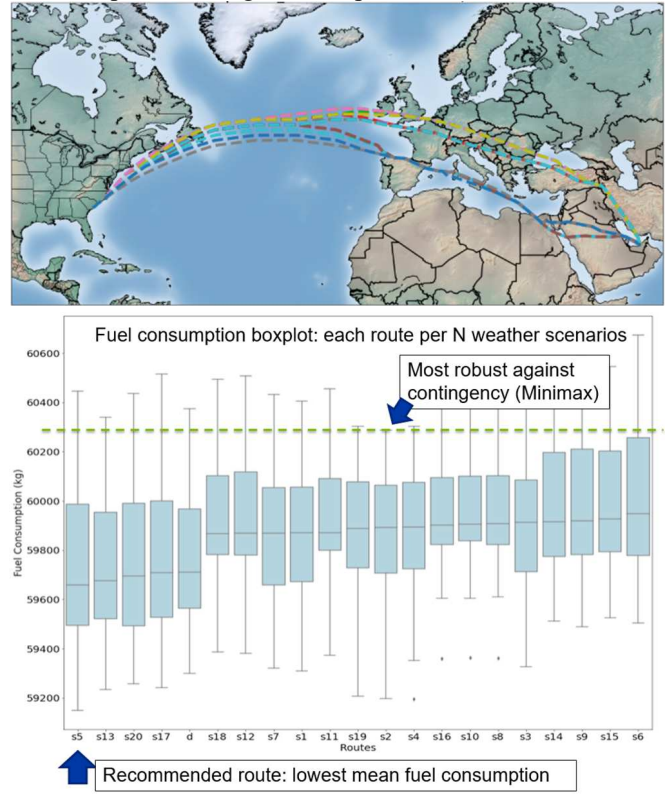

Fig. 11: Stochastic optimization results per flight compared, between the cases with fixed payload and with uncertain payload.

## VIII. STOCHASTIC SCHEDULE PLANNING

We extended the optimization idea to work on whole schedules, including a program of many flights. This problem is extremely broad, because it involves the expected demand, the available fleet with different aircraft types, the prices of jet fuel, etc. The most common airline practice is to build a schedule manually, based on their multidisciplinary experiences and adjust iteratively. Although this is effective for most of the airlines, this manual process does not facilitate the quick assessment of alternative scenarios, which would be desirable in a quickly changing socio-economic environment, as it was during the COVID-19 crisis and the recent geopolitical crisis. Thus, probing full schedules with stochastic boundary conditions would help to improve airline profitability.

Our initial approach had fixed flight missions and we varied just the aircraft assignments, with a fleet with different aircraft types, corresponding to different capacities and performance characteristics. Due to the present paper length requirement, we just outline our approach for Stochastic Schedule Planning (SSP), as follows:
1. The airline planner inputs flight missions with origin, destination and time of departure, and the available fleet.
2. SSP system generates multiple schedules that meet the mission combination.

3. Under assumptions on demand and aircraft occupation, SSP generates optimal flight plans and evaluates the operational profit of that schedule.
4. SSP display schedule performance in a visual and succinct approach to the airline planning staff.
5. The airline planning staff choose the schedule that they consider the best.

We tested out this approach for a summer season, where the source of uncertainty being the ensemble weather forecast, evaluating a total of 1080 flights spread in 6 different schedules, and observed a variation of up to 10% in the schedule profit, calculated with the CI-based flight cost.

## IX. FINAL REMARKS

This research work has demonstrated the potential of stochastic flight plan optimization with quantitative benefits on fuel consumption and total flight cost. The uncertain variables explored here were the weather forecast, by means of ensemble forecasts, and payload, by using simple assumptions on the its probabilistic distribution. Among these factors, the weather is the one that contributes with most benefits in the stochastic optimization. It is important to stress that the use of ensemble forecast in these experiments did not cover effects related to severe weather, which may cause capacity reduction and re-routes. The principles of stochastic optimization can be extended to these factors, with appropriate analysis and development, and we believe that this extension would produce larger benefits, as it would be the inclusion of uncertainties on congestion, to name just a few.

By assuming just two factors of uncertainty, given by the weather ensemble forecasts and payload probability distributions, the computational effort required is already massive. If no clever scheme is adopted, the number of core optimization calls is exponential on the number of uncertain variables, thus the full factorial method adopted in this research would become unfeasible just by adding one more uncertain variable. Therefore, more efficient methods of generating combinations of values for the random variables will be required, such as Latin Hypercube Sampling (LHS) or other method with a similar purpose [15].

## X. ACKNOWLEDGEMENTS

Our highest appreciations for: NCEP (NOAA) for their prompt support in answering questions and requests regarding their weather forecast data; Gol Linhas Aéreas Inteligentes for providing analysis data; and Etihad for their enthusiasm in testing our prototype and providing feedback; all extremely important elements in this research.

## XI. REFERENCES

[1] J. C. Spall, Introduction to Stochastic Search and Optimization, Wiley, 2003.

[2] *14 CFR: § 91.167 Fuel requirements for flight in IFR conditions..*

[3] L. J. Bailey, Í. Romani de Oliveira and J. Fregnani, "Vertical flightpath optimization". US Patent 11,094,206 B2, 17 Aug 2021.

[4] A. Olivares, M. Soler and E. Staffetti, "Multiphase Mixed-Integer Optimal Control applied to 4D Trajectory Planning in Air Traffic Management," in *Proceedings of: International Conference on Application and Theory of Automation in Command and Control Systems (ATACCS)*, Naples, 2013.

[5] A. Jouglet and J. Carlier, "Dominance rules in combinatorial optimization problems," *European Journal of Operations Research,* vol. 212, pp. 433-444, 2011.

[6] "Route Availability Document," [Online]. Available: https://www.nm.eurocontrol.int/RAD/. [Accessed 11 1 2023].

[7] "Ensemble Forecasting," [Online]. Available: https://en.wikipedia.org/wiki/Ensemble_forecasting. [Accessed 04 01 2023].

[8] Z. Toth and E. Kalnay, "Ensemble Forecasting at NCEP and the Breeding Method," *Monthly Weather Review,* vol. 125, no. 12, p. 3297–3319, 1997.

[9] F. Molteni, R. Buizza, T. Palmer and T. Petroliagis, "The ECMWF Ensemble Prediction System: Methodology and validation," *Quarterly Journal of the Royal Meteorological Society,* vol. 122, no. 529, p. 73–119, 1996.

[10] "The Global Ensemble Forecast System," [Online]. Available: https://www.ncei.noaa.gov/products/weather-climate-models/global-ensemble-forecast. [Accessed 04 01 2023].

[11] "Global Data Assimilation System," [Online]. Available: https://www.ncei.noaa.gov/products/weather-climate-models/global-data-assimilation. [Accessed 12 1 2023].

[12] B. Roberson, "Fuel conservation strategies: Cost Index Explained," [Online]. Available: https://www.boeing.com/commercial/aeromagazine/articles/qtr_2_07/article_05_1.html. [Accessed 13 1 2023].

[13] Etihad, "2021 Rome ecoFlight Highlights," [Online]. Available: https://www.etihad.com/en/news/2021-rome-ecoflight-highlights. [Accessed 16 1 2023].

[14] IATA, "Airline and location code search," [Online]. Available: https://www.iata.org/en/publications/directories/code-search/. [Accessed 17 1 2023].

[15] M. D. Shields and J. Zhang, "The generalization of Latin hypercube sampling," *Reliability Engineering and System Safety,* vol. 148, pp. 96-108, 2016.